\documentclass[twocolumn,prb,nobibnotes,altaffilletter,amsmath,amssymb,amsfonts]{revtex4}
\usepackage[latin1]{inputenc}
\usepackage{graphicx}
\usepackage{amsmath}
\usepackage{latexsym}
\usepackage{epsfig}
\usepackage{color}
\usepackage{verbatim}
\usepackage[breaklinks=false]{hyperref}

\usepackage{amssymb, dsfont}

\newcommand{\f}{{\mathbf f}}

\newcommand{\beq}{\begin{equation}}
\newcommand{\eeq}{\end{equation}}
\newcommand{\bea}{\begin{eqnarray}}
\newcommand{\eea}{\end{eqnarray}}

\begin{document}

\begin{@twocolumnfalse}
\begin{center}
   \mbox{
{
\color{blue} \href{http://dx.doi.org/10.1103/PhysRevLett.113.238303}{http://dx.doi.org/10.1103/PhysRevLett.113.238303}
}
    }
\end{center}
  \end{@twocolumnfalse}


\title{Generalized energy equipartition in harmonic oscillators driven by active baths}

\author{Claudio Maggi$^1$}
\email{claudio.maggi@roma1.infn.it}
\author{Matteo Paoluzzi$^2$}
\author{Nicola Pellicciotta$^1$}
\author{Alessia Lepore$^1$}
\author{Luca Angelani$^2$}
\author{Roberto Di Leonardo$^{1,2}$}

\affiliation{ $^1$Dipartimento di Fisica, Universit\`a di Roma ``Sapienza'', I-00185, Roma, Italy }
\affiliation{ $^2$CNR-IPCF, UOS Roma, Dipartimento di Fisica Universit\`a
Sapienza, I-00185 Roma, Italy}
\date{\today}

\begin{abstract}

We study experimentally and numerically the dynamics of colloidal beads confined by a harmonic potential in a bath of swimming \textit{E. coli} bacteria. The resulting dynamics is well approximated by a Langevin equation for an overdamped oscillator driven by the combination of a white thermal noise and an exponentially correlated active noise. This scenario leads to a simple generalization of the equipartition theorem resulting in the coexistence of two different effective temperatures that govern dynamics along the flat and the curved directions in the potential landscape.

\end{abstract}

\maketitle

\textit{Introduction}- 
A remarkable result of equilibrium statistical mechanics is the theorem of energy equipartition. In its simplest form, the theorem states that each quadratic term in the Hamiltonian contributes with the same amount of energy $k_B T/2$ to the average energy of the system~\cite{Huang}. In the case of a harmonic oscillator
this applies to both kinetic and potential energies. In the colloidal realm, particle motions are strongly overdamped and velocity fluctuates on a timescales that is often hardly accessible~\cite{Inertia}. However the value of kinetic energy imposed by the equipartition theorem is reflected in a diffusion coefficient that is proportional to the mean squared velocity~\cite{Risken}: $D_T=\mu k_B T$.
Therefore, for a colloidal harmonic oscillator the equipartition theorem establishes a link between the thermal diffusion constant $D_T$ and the average potential energy $U=D_T/2\mu$.
Out of equilibrium system are frequently found in nature and the search for generalized equipartition laws constitutes a very interesting and hot topic~\cite{To,Conti}.
In particular there is a growing family of off-equilibrium, active colloidal particles that are able to harness some form of locally stored energy to self propel in persistent random walks \cite{poon, cates}.  An interesting example is provided by passive colloidal tracers suspended in active baths of swimming bacteria. Over time scales that are larger than the persistence time of active forces,  those particle display a diffusive behaviour with a diffusivity $D^*$ that can be order of magnitudes larger than the thermal counterpart $D_T$~\cite{SinSph}. It is found that, whenever the external potential changes smoothly on the characteristic length-scale of the persistent motion, the system is well described by a quasi-Boltzmann distribution with an effective temperature given by $k_B T_{\mathrm{eff}}=D^*/\mu$~\cite{JanusGrav,Centri}. In this limiting case the equipartition theorem is recovered in its original form, being a straightforward consequence of Boltzmann statistics.  However, when the external potential does not meet these requirements, Boltzmann statistics breaks down~\cite{CatesEPL,CatesRep} and an equilibrium-like picture with one single effective temperature fails.  
This is particularly evident in the case of rectification effects, as those  investigated in Ref.s~\cite{RatchetSim,Ratchet,Shuttle}. In these works it has been shown that a bacterial bath can spontaneously induce the unidirectional motion of nano-fabricated asymmetric objects. Similarly, micro-fabricated structures can rectify the motion of motile bacteria and accumulate them in specific spatial regions~\cite{Funnels}. Moreover passive colloidal tracers can be delivered onto target sites by the rectification of fluctuating forces from a bacterial bath~\cite{Struct,StructSim}. Failure of Boltzmann statistics also leads to novel non-equilibrium effects such as the emergence of effective attraction in presence of purely repulsive potentials~\cite{ActDep,Ivo}. In this context, a simple generalization of equipartition could seem unlikely.

In this Letter we demonstrate that, in the case of active harmonic oscillators, the average value of potential energy is still linked to the diffusivity by a simple generalization of the equipartition theorem. As a consequence, the effective temperature that associated with the potential energy is always lower than the one obtained from the free diffusion coefficient.
We investigate experimentally and numerically the dynamics of colloidal beads, subject to a harmonic potential, suspended in a bath of swimming  \textit{E. coli} cells. The elastic force field is obtained experimentally by placing the micro-spheres in a cylindrical microcapillary. Sedimenting colloids fluctuate near the bottom of the capillary where they experience a near-perfect harmonic potential.

\begin{figure}
\begin{center}
\includegraphics[width=8.5cm]{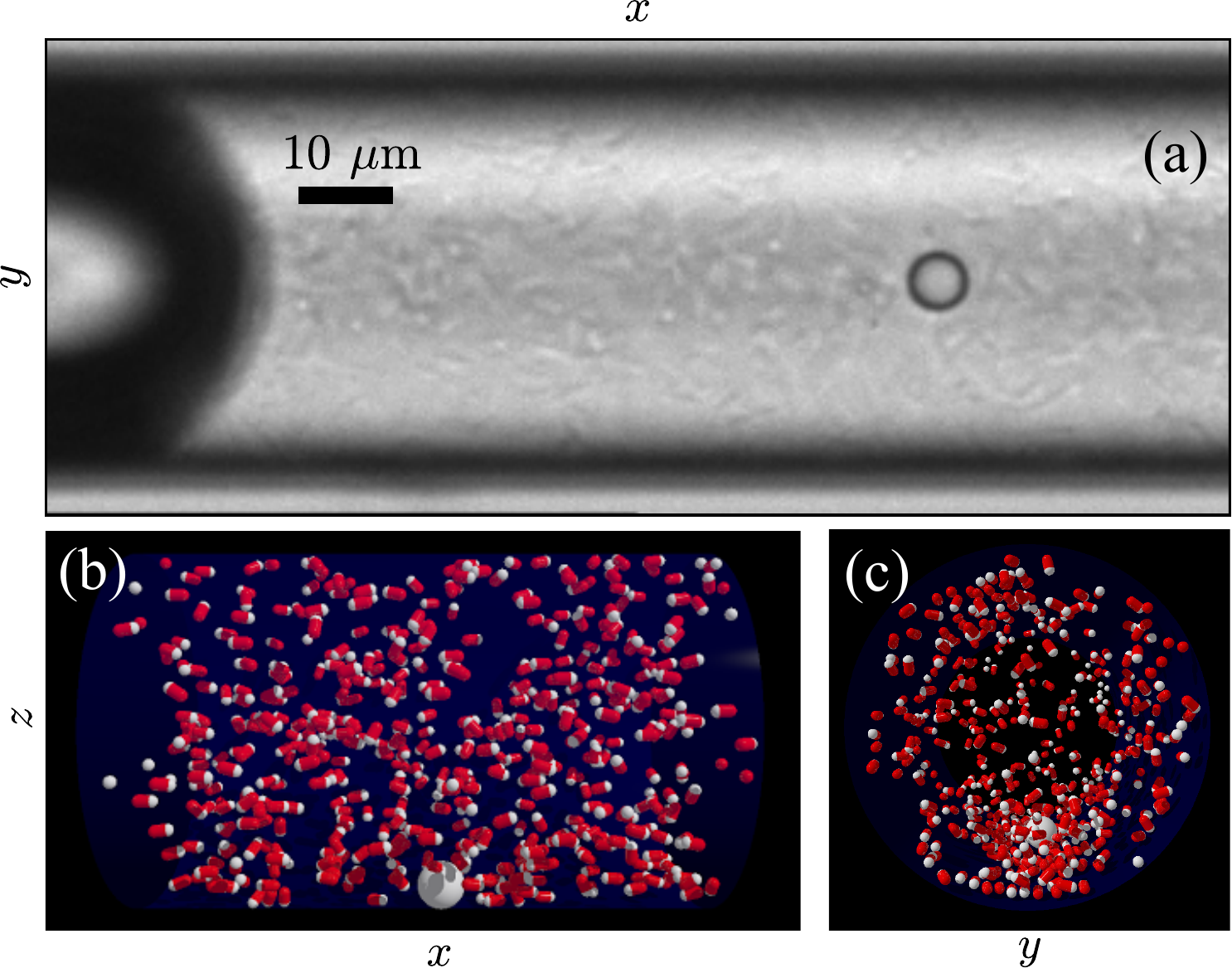}
\caption{ 
(\textbf{a}) A $3.5 \, \mathrm{\mu m}$ radius silica bead is suspended in a bath of motile \textit{E. coli}
bacteria filling a $25 \, \mathrm{\mu m}$ radius capillary glass tube. 
(\textbf{b}),(\textbf{c}) Snapshots from the numerical simulation.
}
\label{fig:f0}
\end{center}
\end{figure}

\textit{Experiment}- 
Motile \textit{E. coli} cells are prepared following the protocol described in~\cite{Supp}.
Silica beads of radius $a=3.5 \, \mathrm{\mu m}$ are first diluted in
deionized water and then mixed with bacteria directly on a glass
slide.  The final bacteria density is estimated to be $\sim 10^{10}$ cells/ml. 
The bacteria-colloids solution is loaded in a microcapillary glass tube (Vitrocom) of 
internal radius $R= 25 \, \mathrm{\mu m}$  by capillarity.
The sample is left open for few minutes before sealing with index matching oil.
This procedure results in the formation of two air bubbles at the edges of
the capillary tube as shown in Fig.~\ref{fig:f0}(a). 
{Residual distortions due to the internal glass/water interface have a negligible effect as shown by the absence of anisotropies in both particle shape and diffusion 
in absence of bacteria (see~\cite{Supp}).}

\begin{figure*}
\begin{center}
\includegraphics[width=18.0cm]{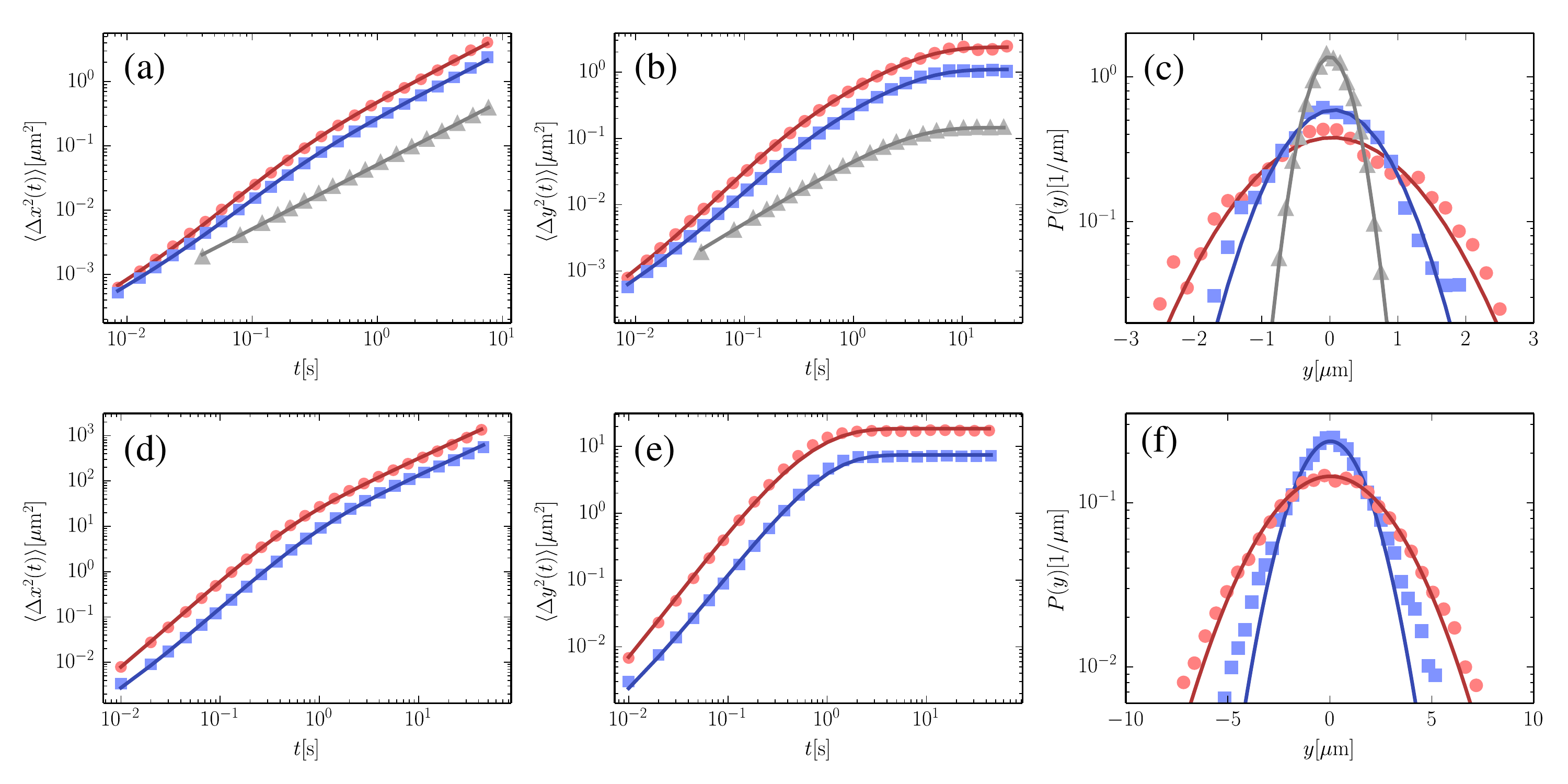}
\caption{(Top panels-Experimental data) 
(\textbf{a}) 
MSD along the capillary axis of two colloids located at $30$ ($\circ$) and at $230 \, \mathrm{\mu m}$ ($\Box$)
from the air bubble (oxygen reservoir), ($\bigtriangleup$) MSD of 
the colloid in absence of bacteria~\cite{Supp}. The full lines are fits with Eq.~(\ref{eq:msd}). 
(\textbf{b}) 
MSD of the same particles in (a) along an orthogonal direction to the capillary axis.  Full lines are fits with Eq.~(\ref{eq:spec}).
(\textbf{c}) 
Probability distributions of $y$ of the same two colloids in (a)
and (b), the full lines are fits with a Gaussian distribution.
(Bottom panels-Simulation data) 
(\textbf{d}), (\textbf{e})  and (\textbf{f}) are the same as (\textbf{a}), (\textbf{b})  and (\textbf{c}) for two simulated colloids
in a bacterial bath with varying average swimming speed $30 \, \mathrm{\mu m /s}$ 
($\circ$) and $15 \, \mathrm{\mu m /s}$ ($\Box$). 
}
\label{fig:f1}
\end{center}
\end{figure*}

{ Colloidal beads sediment at the bottom of the capillary and align along the tube axis with an average distance of about 20 $\mu$m. 
We collect bright field images using a $20\times$, NA 0.25 microscope objective. 
After background subtraction and thresholding we obtain particle trajectories by center of mass tracking. We report data for 10 beads that were simultaneously tracked for 100 s at a rate of 100 frames/s. The beads span a capillary length of approximately 250 $\mu$m probing a local environment characterised by a bacterial activity that decreases as the distance from the trapped air bubble increases.}


{\it Simulation}- The numerical simulations are performed by
considering spherical colloidal particles of radius $a$ immersed in a
bath of self-propelling dumbells following a ``run and tumble" dynamics.
Both particles and bacteria are
confined in a cylindrical volume as shown in Fig.~\ref{fig:f0}(b) and (c). 
All interactions are modelled by repulsive steric forces.
In addition particles experience a gravitational force
$f_z = -mg$ due to gravity where $m$ is the buoyant mass of the colloidal particle
and $g$ is the acceleration due to gravity. 
We include Brownian motion only for particles dynamics and neglect hydrodynamic interactions~\cite{RatchetSim,ActDep}. A detailed description of the simulation can be found in~\cite{Supp}. 


\textit{Results}-
The mean-squared displacement  $\langle \Delta x^2(t) \rangle$ (MSD) along the capillary axis 
is shown in Fig.~\ref{fig:f1}(a) for two beads located at about $30 \, \mathrm{\mu m}$ and at $230 \, \mathrm{\mu m}$ from the edge of air bubble. 
{Both particles show an MSD characterized by a super-diffusive regime at short times
followed by a diffusive dynamics at longer time-scales.
This is in qualitative agreement with the results of previous experiments performed on flat surfaces~\cite{SinSph,Chantal,Struct}.} 
Along the $y$-axis the particle motion is constrained by the curved capillary surface and the MSD saturates at a constant value (Fig.~\ref{fig:f1}(b)). 
The MSD along both axes decreases when moving away from the air bubble, due to a decrease of bacterial activity, as discussed more in detail in the following. Fig.~\ref{fig:f1}(c) shows the probability distribution $P(y)$ together with the best Gaussian fits.

A very similar behaviour is observed in simulations where the diffusivity of colloids is tuned by varying the average swimming speed of  bacteria 
(see Fig.~\ref{fig:f1}(d), (e) and (f)).
In particular when bacteria are faster the diffusivity of the particles increases but also
the transition between the ballistic and the diffusive regime shifts to a shorter 
time-scale.

\textit{Modelling and Discussion}- 
To model the dynamics of the microbeads both in experiment and simulations we assume that
the motion of the particles is mostly constrained on the capillary surface. 
In this case the $z$-coordinate of the center of mass of the 
colloidal particle is directly given by 
$z = -\sqrt{\rho^2-y^2} \sim-\rho+y^2/2\rho$ with $\rho=R-a$. 
The force acting along the $y$ axis is therefore computed as 
$f_y = -m g \, \partial z / \partial y \sim- m g y/\rho$. The resulting force field is then well approximated by an 
elastic force acting along the $y$ axis with a spring constant $k$ defined by 
$k=mg/\rho$.
The cylindrical geometry of the capillary results in $f_x = 0$.
In addition to the deterministic force $\f$ the colloids are subject 
to the thermal fluctuations and to the interactions with the swimming bacteria. 
To account for these we model the dynamics of the beads with the following stochastic 
differential equation

\begin{equation} \label{eq:dyneq}
\dot{\mathbf{r}} = \mu\mathbf{f}(\mathbf{r}) + \boldsymbol{\eta}^T + \boldsymbol{\eta}^A
\end{equation}

\noindent where 
$\mathbf{r}(t)=(x(t),y(t))$, 
$\mathbf{f}=(0,-k \, y)$ and
$\mu$ is the mobility of the colloidal particle. 
We assume that the noise term can be split into two independent components: the standard Langevin thermal noise
$\boldsymbol{\eta}^T$  with  $\langle \eta_{\alpha}^T(t) \eta_{\beta}^T(t') \rangle = 2 D_T \delta_{\alpha \beta} \delta (t-t')$
and $D_T = \mu \, k_B T$;  an active noise $\boldsymbol{\eta}^A$ that is exponentially time-correlated~\cite{StructSim}
$\langle \eta_{\alpha}^A(t) \eta_{\beta}^A(t') \rangle = D_A \delta_{\alpha \beta} \exp(-|t-t'|/\tau)/\tau$, where $\alpha,\beta$ represent individual Cartesian components.
From Eq.~(\ref{eq:dyneq}) we can compute the MSD along $x$:

\begin{equation} \label{eq:msd}
\langle \Delta x^2(t) \rangle = 2 D_T t + 2 D_A \left[t - \tau (1-e^{-t/\tau})\right]
\end{equation}

\noindent and the MSD of $y(t)$:

\begin{eqnarray}
\langle \Delta y^2(t) \rangle &=&  
\frac{2 D_T}{\mu k} (1-e^{-\mu k t}) \nonumber
\\ &+& \frac{2 D_A}{\mu k}\ \frac{1-e^{-\mu k t} - \mu k \tau (1-e^{-t/\tau})}{1-(\mu k \tau)^2} \label{eq:spec}
\end{eqnarray}

Equations (\ref{eq:msd}) and (\ref{eq:spec}) provide an excellent fit to the MSD along both $x$ and $y$ (Fig.s~\ref{fig:f1}(a) and (b)). 
Along both axes the parameter $D_A$ shows a clear dependence on the average position of the particle $\langle x \rangle$ with respect to the edge of air bubble.
$D_A$ is found to decrease from $0.31$ to $0.14 ~\mathrm{\mu m^2/s}$
upon increasing $\langle x \rangle$ by few hundreds microns.
This suggests that bacterial motility depends on the concentration of 
oxygen that is progressively consumed by bacteria along the capillary~\cite{oxy}. Differently the fitting parameters $D_T$, $\tau$ and $\mu k$ do not show any clear dependence on the distance from  the air bubble. The obtained averages
over all particles  are $\tau = 0.093(0.015)$~s,
$D_T$ = 0.030(0.002)~$\mu$m $^2$/s  along $x$ and 
$\tau = 0.097 (0.023)$~s,
$D_T = 0.046 (0.013)~\mathrm{\mu m^2/s}$,
$\mu k = 0.289 (0.067)~\mathrm{s^{-1}}$ along $y$, where standard deviations 
are shown in brackets.
MSD in absence of bacteria are reported in Fig.~\ref{fig:f1}. 
The corresponding fitting parameters are $\mu k = 0.36$ s$^{-1}$ 
and $D_T=0.025$ and $0.026 \mathrm{\mu m^2/s}$ along $x$ and $y$ respectively. 
Those values are compatible with those found in presence of bacteria, although correlations between fitting parameters result in larger uncertainties in $D_T$ along $y$.
The values of $D_T$, with or without bacteria, 
are about a factor two smaller than the bulk value which can be attributed increased drag due to wall effects~\cite{walldrag,Supp}.
We can now estimate the sedimentation length of our colloids in the bacterial bath as $(D_T+D_A)/\mu m g\sim 40$ nm which validates our initial assumption of a colloidal motion that is mostly restricted at the capillary surface.
{Previous studies of bacterial swimming in the presence of confining walls have evidenced the possibility of complex swimming patterns that could possibly result in anisotropies in the bacterial bath~\cite{circ, pre}. Such an anisotropy, if present, should be reflected in a corresponding asymmetry in the motion of the colloidal tracers along $x$ and $y$.
However, we do not observe any systematic deviation between $D_A$ and $\tau$ when fitted independently to the $x$ and $y$ components of MSD.}

The MSD along $x$ and $y$ of the beads from simulations can be fitted with the same 
Equations (\ref{eq:msd}) and (\ref{eq:spec}) where this time 
$D_A$ and $\tau$ are the only free parameters. 
As seen in Fig.~\ref{fig:f1}(d) and (e) these
functions fit very well the simulation data. 
By fitting the MSD along $x$ with the free parameters $D_A$ and $\tau$ we find that $D_A$ 
grows continuously from $6.8$ to $16.0$ $\mathrm{\mu m^2/s}$ as we increase the average speed from $15$ to $30$ $\mathrm{\mu m/s}$.
The parameter $\tau$ shows also a marked change upon 
changing the average speed going from $\tau=0.44$ to $0.23$ s.

The $\langle \Delta y^2(t) \rangle $ from simulations 
is fitted with Eq.~(\ref{eq:spec}) with the same free parameters $D_A$ and $\tau$ giving
$D_A$ growing from $6.7 $ to $14.0 \, \mathrm{\mu m^2/s}$, that is almost 
identical to the one found from the fitting of $\langle \Delta x^2(t) \rangle $. 
Also the $\tau$ found from the fit of $\langle \Delta y^2(t) \rangle $ 
is very close to the one found from the MSD decreasing from 
$\sim 0.4$ to $0.2$ s upon increasing $V$. 
It has to be noted that simulations are in qualitative agreement with the experiments although a quantitative comparison shows that both $\tau$ and $D_A$ result considerably larger in simulations than in experiments. 

\begin{figure}
\begin{center}
\includegraphics[width=8.5cm]{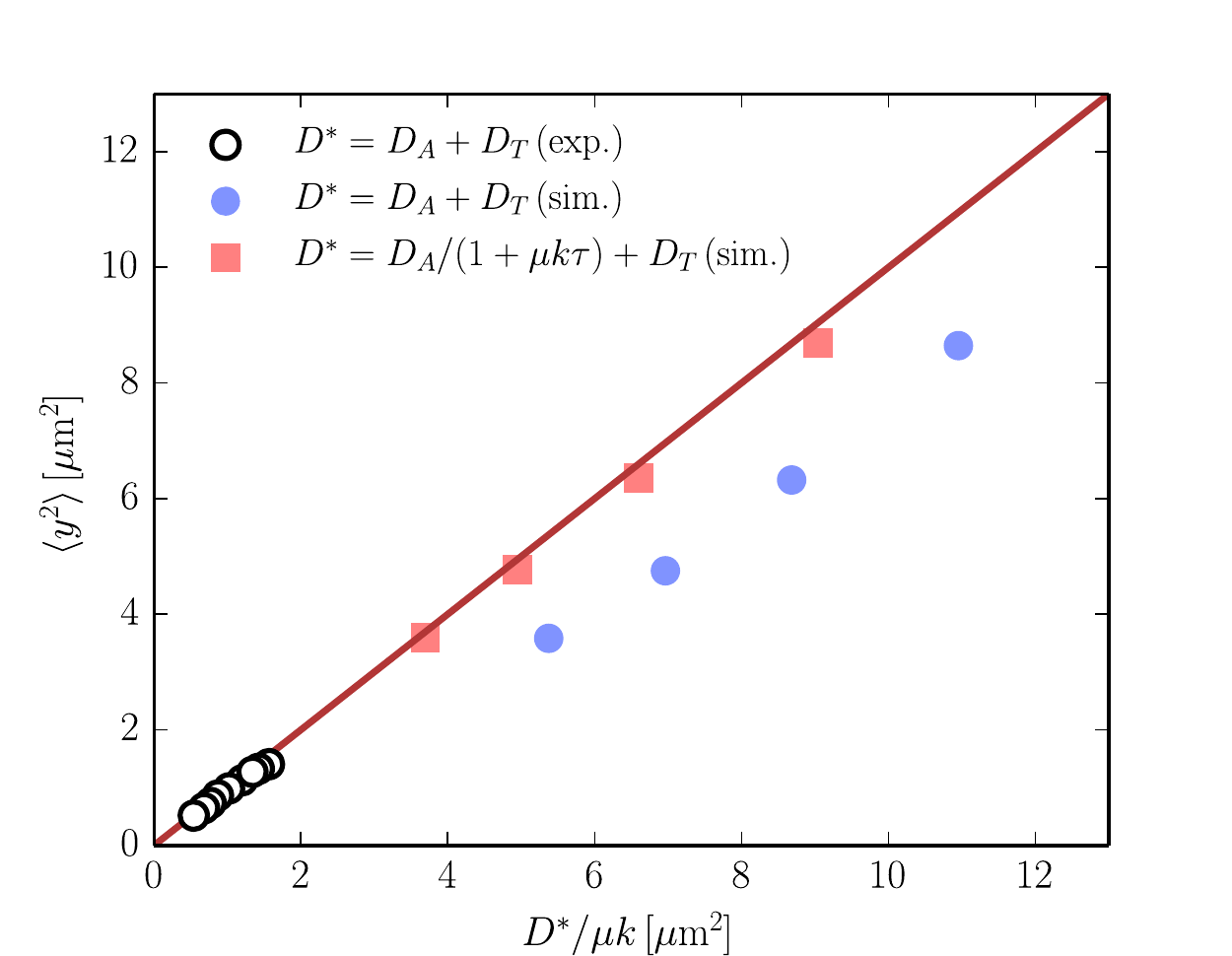}
\caption{
Generalized equipartition plot. The variance of particle fluctuations along the $y$-coordinated is proportional 
to a weighted sum of thermal and active diffusivities Eq.~(\ref{eq:equi}).
}
\label{fig:f3}
\end{center}
\end{figure}

It is however clear that the model of Eq.~(\ref{eq:dyneq}) can be used to fit both numerical and experimental curves and that this allows us to make a precise statement on how to generalise equipartition of energy for active particle systems in harmonic potentials. At equilibrium, when only thermal noise is present, the average potential energy of the particle $U  = k \langle y^2 \rangle/2$ is simply given by the equipartition theorem $U = D_T/2\mu$. When active forces are introduced, they will add an extra contribution to $U$ that can be obtained taking the limit of (\ref{eq:spec}):

\begin{equation} \label{eq:equi}
U=
\frac{1}{2} k \lim_{t \to \infty}  \langle \Delta y^2(t)\rangle/2 = 
\frac{D_T}{2\mu}+\frac{D_A}{2\mu}\frac{1} {1+\mu k \tau}
\end{equation} 

It is worth noting that, even within the very general premises of exponentially correlated noise, the expression for the average potential energy retains a form that is very close to the equilibrium equipartition result, with the only difference that the contribution from the active noise is reduced by a factor $1+\mu k \tau$. When the persistence time $\tau$ is much shorter than the relaxation time in the potential well $1/\mu k$, we recover the equilibrium form. This result is direct consequence of the fact that when $\tau$ is much shorter than any other time scale in the problem, the active noise is practically white and Boltzmann statistics holds with the (unique) effective temperature $k_B T_\mathrm{eff} = (D_A+D_T)/\mu$. However if $\tau \mu k\sim 1$, even when the stationary distribution deviates strongly from the Boltzmann (as in run and tumble dynamics~\cite{CatesEPL}), the average potential energy will be given by the simple formula (\ref{eq:equi}).

We now discuss this generalized equipartition formula 
in both experiments and simulation. 
We plot the variance $\langle y^2 \rangle$ in Fig.~\ref{fig:f3} . 
Fig.~\ref{fig:f3} shows that the experimental $\langle y^2 \rangle$ 
is close to the straight line when plotted as a function of 
$(D_A+D_T)/\mu k$ indicating only weak deviation from the unique 
$T_\mathrm{eff}$ regime. This is consistent with the experimental
$\mu k \, \tau \approx 0.028 \, ( \ll 1)$. 
On the other hand in simulations $\mu k \tau$ ranges approximately 
from 0.47 to 0.23. As shown in Fig.~\ref{fig:f3} 
we observe $\langle y^2 \rangle$ deviating considerably from the straight line
when plotted as a function of $(D_A+D_T)/\mu k$, while when plotted as a function of Eq.~(\ref{eq:equi}) we see a substantial agreement with the straight line.

We remark that all these considerations are restricted to the second moment of fluctuations
(i.e. $\langle y^2 \rangle$), and remain valid as long as the active noise is exponentially correlated, whatever is the static noise distribution.
In our specific case, we have  empirically found that the probability distribution $P(y)$ is well approximated by a Gaussian. 
This implies that $P(y) \sim \exp[-  k y^2 / 2 k_B T_{\mathrm{eff}} ]$ and all the static properties of the colloids in the active bath under the influence of the harmonic potential can be predicted by setting $k_B T_\mathrm{eff} = D_A/[\mu (1+\mu k \tau)]+D_T/\mu$. This effective temperature is however different from the effective temperature governing long time diffusion along the flat direction $x$ which is given by $(D_A+D_T)/\mu$. In a way  the system behaves like an equilibrium system whose free diffusivity and potential energy are governed by two different effective temperatures, the latter being a function of the curvature $k$ of the external potential.

\textit{Conclusions}- 
We have investigated,
experimentally and numerically, the possibility of generalizing energy equipartition to out of equilibrium systems
consisting of colloidal particles that are subject to both a harmonic potential and the interactions with a bath of swimming bacteria.
We found that the system obeys a modified energy equipartition law. A harmonic degree of freedom contributes an average potential energy that takes the equilibrium form for small curvatures and decreases when the relaxation time in the harmonic well starts to be comparable to the persistence time of active forces.
Based on these observations we expect that using a different type of self propelled colloids, i.e. Janus particles, one could have direct experimental access to higher $\tau$ values and observe the predicted strong deviations from equilibrium equipartition. 

The research leading to these results has received funding from the European Research Council under the European Union's Seventh Framework Programme (FP7/2007-2013) / ERC grant agreement n$^\circ$ 307940. We also acknowledge funding from  MIUR-FIRB project No. RBFR08WDBE.



\begin{thebibliography}{99}

\bibitem{Huang}
K. Huang, \textit{Statistical Mechanics} (John Wiley, New York, 1987).

\bibitem{Inertia}
B. Lukic, S. Jeney, C. Tischer, et al
Phys. Rev. Lett. \textbf{95}, 160601 (2005)

\bibitem{Risken}
H. Risken: \textit{The Fokker-Planck Equation. Methods of Solution and Applications} Springer (1984)

\bibitem{To} 
K. To, Phys. Rev. E \textbf{89} 062111 (2014)

\bibitem{Conti} 
L. Conti, P. De Gregorio, G. Karapetyan J. Stat. Mech.  P12003 (2013)

\bibitem{poon}
W. K. Poon,  
Proceedings of the International School of Physics ``Enrico Fermi",
Course CLXXXIV ``Physics of Complex Colloids", 
edited by C. Bechinger, F. Sciortino and P. Ziherl (IOS, Amsterdam; SIF, Bologna)

\bibitem{cates}
M. E. Cates, Reports on Progress in Physics, \textbf{75}, 042601 (2012).

\bibitem{SinSph} X.L.~Wu and A.~Libchaber, 
Phys. Rev.  Lett.  \textbf{84}, 3017 (2000). 

\bibitem{JanusGrav} 
J. Palacci, C. Cottin-Bizonne, C. Ybert,  L. Bocquet, Phys. Rev. Lett. \textbf{105} 088304 (2010)

\bibitem{Centri}
C. Maggi, A. Lepore, J. Solari, Soft Matter, 9, 10885-10890 (2013)

\bibitem{CatesEPL}
J. Tailleur and M. E. Cates, EPL, 86, 60002 (2009)

\bibitem{CatesRep}
M. E. Cates, Rep. Prog. Phys.,  75, 042601 (2012)

\bibitem{RatchetSim} L. Angelani, R. Di Leonardo, G. Ruocco,
Phys. Rev. Lett., \textbf{102}, 048104, (2009) 

\bibitem{Ratchet} R. Di Leonardo et al.,
PNAS, \textbf{107}, 9541, (2010) 

\bibitem{Shuttle} L. Angelani, R. Di Leonardo,
New Journal of Physics, \textbf{12}, 113017, (2010) 

\bibitem{Funnels}  P. Galajda et al., J. Bacteriol. 189, 8704 (2007).

\bibitem{Struct} N. Koumakis, A. Lepore, C. Maggi, R. Di Leonardo,
Nature Communications, \textbf{4}, 2588, (2013).

\bibitem{StructSim} N. Koumakis, C. Maggi, R. Di Leonardo,
DOI: 10.1039/C4SM00665H (2014) 

\bibitem{ActDep} L. Angelani, C. Maggi, M. L. Bernardini, A. Rizzo, R. Di Leonardo,
Phys. Rev. Lett., \textbf{107}, 138302, (2011).

\bibitem{Ivo}
Ivo Buttinoni, J. Bialke, F. Kummel, et al. 
Phys. Rev. Lett. \textbf{110}, 238301 (2013)

\bibitem{Supp} Supplemental Material  which includes Refs.~\cite{brock,E_coli,coli1,coli2,Martinez12,kim,numrec}

\bibitem{Chantal} C. Valeriani, M. Li, J. Novosel et al.
Soft Matter, \textbf{7}, 5228 (2011).

\bibitem{walldrag}  
E. Schaffer, S. F. Norrelykke, and J. Howard, Langmuir
\textbf{23}, 3654 (2007)

\bibitem{circ}
S. van Teeffelen, U. Zimmermann, H. Löwen, Soft Matter \textbf{5}, 4510-4519 (2009)

\bibitem{pre}
P. K. Radtke and L. Schimansky-Geier Phys. Rev. E \textbf{85}, 051110 (2012).

\bibitem{oxy} C. Douarche, A. Buguin, H. Salman et al. Phys. Rev. Lett. \textbf{102},
198101 (2009).


\bibitem{brock} M. T. Madigan et al.,
Brock Biology of Microorganisms 13th edition, Prentice Hall (2008)

\bibitem{E_coli}
H.C.~Berg,
{\it E. Coli in Motion} 
(Springer-Verlag, New York, 2004).

\bibitem{coli1}
S.~ Chattopadhyay, R.~Moldovan, C.~Yeung, and X.L.~Wu,
Proc. Natl. Acad. Sci. U.S.A.  \textbf{103}, 13712 (2006).

\bibitem{coli2}
N.C.~Darnton, L.~Turner, S.~Rojevsky, and H. C.~Berg, 
J. Bacteriol., {\bf 189}, 1756 (2007).

\bibitem{Martinez12} V. A. Martinez, R. Besseling, O. A. Croze, J. Tailleur, M. Reufer, J. Schwartz-Linek, L. G. Wilson, M. A. Bees, and W. C. K. Poon, Biophys. J. {\bf{103}}, 1637 (2012).

\bibitem{kim} S. Kim, S. Karrila, Microhydrodynamics, Dover, New York
  (2005).
  
\bibitem{numrec}  W.H.~Press, W.T.~Vetterling, S.A.~Teukolsky, and
  B.P.~Flannery,  {\it Numerical Recipes in C}  (Cambridge University Press,
  Cambridge, England, 1992), 2nd ed.


\end{thebibliography}
\end{document}